\documentclass[10pt]{article}
\begin{document}
\title{Observables in terms of connection and curvature variables for Einstein's equations with two commuting Killing vectors\footnote{appeared as: Kordas P. 2015 Observables
in terms of connection and curvature variables
for Einstein’s equations with two commuting
Killing vectors. Proc. R. Soc. A 471: 20150350.
http://dx.doi.org/10.1098/rspa.2015.0350}}
\author{P Kordas\\35 Square Marie-Louise\\1000 Bruxelles\\Belgium\\email: panayiotis.kordas@physics.org}
% Rmove command to get current date

\maketitle
% Remove command to get current date 
\begin{abstract}
Einstein's equations with two commuting Killing vectors and the associated Lax pair are considered. The equations for the  connection $A(\varsigma, \eta, \gamma)=\Psi_{,\gamma}\Psi^{-1}$, where $\gamma$ the variable spectral parameter are considered. A transition matrix ${\cal T}= A(\varsigma, \eta, \gamma)A^{-1}(\xi, \eta, \gamma)$ for $A$ is defined relating $A$ at ingoing and outgoing light cones. It is shown that it satisfies equations familiar from integrable pde's theory. A transition matrix on $\varsigma={\mbox constant}$ is defined in an analogous manner.     

These transition matrices allow us to obtain a hierarchy of integrals of motion with respect to time, purely in terms of the trace of a function of the connections $g_{,\varsigma}g^{-1}$ and $g_{,\eta}g^{-1}$. Furthermore a hierarchy of integrals of motion in terms of the curvature variable $B=A_{,\gamma}A^{-1}$, involving the commutator $[A(1), A(-1)]$,  is obtained. 

We interpret the inhomogeneous wave equation that governs $\sigma=ln N$, $N$ the lapse, as a Klein-Gordon equation, a dispersion relation relating energy and momentum density, based on the first connection observable and hence this first observable corresponds to mass. The corresponding quantum operators are $\frac{\partial}{\partial t}$, $\frac{\partial}{\partial z}$ and this means that the full Poincare group is at our disposal.
\end{abstract}

\thanks{Keywords: General Relativity, Integrable Systems, Gravitation, Quantum Field Theory, Klein-
Gordon equation}

\section{Introduction}
\label{sec:Intro}
 The general relativistic equations in the presence of two commuting Killing vectors have received a great amount of interest over the years. The first important contribution, relevant to the considerations here, was the proof in \cite{Geroch} that there exists an infinite hierarchy of solutions which can be mapped to one another via transformations of what we now call the Geroch group. 
Later it was shown \cite{BelinskyZakharov1, Maison, Neugebauer, HauserErnst, Harrison, HKX} by different authors and in differing approaches, that the field equations (\ref{eq:field equations g}) are integrable (in the sense understood in the inverse scattering field) and a variety of solutions was obtained and analyzed. These results are presented and reviewed in \cite{BelinskyVerdaguer}. Relatively more recently techniques of quantum inverse scattering were used\cite{KorotkinNicolai} to quantize appropriate functions of the gravitational field appearing in (\ref{eq:field equations g}) with the emphasis on the metric.  The main motivation of the present paper is to present an approach for future work in terms of connections which are appropriate variables for Gravity.  

 The organization of the present work is as follows. In section (\ref{sec:BackgroundForEinsteinSEquationsInTheBelinskyZakharovFormalism}) we introduce the equations and briefly present the solitonic technique. Further in section (\ref{sec: spectral current}) 
$A(\gamma)$ is defined. Transition matrices for $A$ relating it at points on ingoing and outgoing null coordinates is defined, and it is proved that it obeys equations similar to the ones of other integrable pde's, a result absent from the literature thus far, for connection variables. This presents the possibility of obtaining integrals of motion, from the trace of the derivative of appropriate combinations of the two transition matrices,
in terms of classical connections of General Relativity. These transition matrices first appeared in \cite{kordasarx}, where it is also noted that the possibility of observables in terms of connections can be attained. These may be appropriate for quantization as they are the fundamental variables of the classical theory and connections are fundamental also in Quantum Canonical General Relativity \cite{Rovelli, Ashtekar}. It is shown that $Tr(A^{n}(1)A^{-n}(-1))$ is an integral of motion. In section (\ref{sec:B}) we consider the curvature $B(\gamma)=A_{,\gamma}A^{-1}$ and obtain observables in a similar fashion as for the connection, but where now $B(\pm 1)$ involve the commutator $[A(1), A(-1)]$. We discuss these observables in Section (\ref{sec:discussion}), interpreting the first one of the connection observables as mass due to the fact that it appears, as the inhomogeneous term, in the wave equation governing the lapse $N$ which is interpreted as a Klein-Gordon equation. This puts the associated Quantum Field Theory in our hands for quantization.

\section{The Einstein equations with two commuting Killing vectors}
\label{sec:BackgroundForEinsteinSEquationsInTheBelinskyZakharovFormalism}
The metric in the presence of two commuting Killing vectors, and assuming the existence of 2-surfaces orthogonal to the group orbits, is given by \cite{ExactSolutionsBook, BelinskyZakharov1, BelinskyVerdaguer}:
\begin{equation}
\label{eq:metric}
	ds^{2} = f(t,z)(dz^{2}-dt^{2})+g_{ab}(t,z)dx^{a}dx^{b},
\end{equation}
where $g_{ab}(t,z)$ real and symmetric tensor, $a,b= 1, 2$.

Einstein's equations corresponding to this metric are, (in null coordinates $ \varsigma = \frac{1}{2}(z+t)$, $\eta = \frac{1}{2}(z-t)$)
\begin{eqnarray}
\label{eq:field equations g}
(\alpha g_{,\varsigma}g^{-1})_{,\eta}+(\alpha g_{,\eta}g^{-1})_{,\varsigma}=0, \\
\label{eq:field equations f1}
(\ln f)_{,\varsigma}(\ln \alpha)_{,\varsigma} = (\ln \alpha)_{,\varsigma\varsigma}+\frac{1}{4\alpha^{2}} \mbox{Tr} A^{2}(1),\\
\label{eq:field equations f2}
(\ln f)_{,\eta}(\ln \alpha)_{,\eta} = (\ln \alpha)_{,\eta\eta}+\frac{1}{4\alpha^{2}} \mbox{Tr} A^{2}(-1),
\end{eqnarray}
where 
	\begin{equation}
	\label{eq:A(+-1)}
	A(1) = \alpha g_{,\varsigma}g^{-1}, \,\,\,\,\,\,\,\, A(-1) = \alpha g_{,\eta}g^{-1}, 
	\end{equation}
	$ \det g = \alpha^{2} $ and the reason for the labelling $ A(\pm 1) $ will become apparent in the following. Further from the above equations (taking the trace of (\ref{eq:field equations g})) it follows that $ \alpha $ satisfies 
	\begin{equation}
	\label{eq:field equations alpha}
	\alpha_{,\varsigma\eta}=0. \label{eq:alphaequation}
	\end{equation}
Belinsky-Zakharov have shown that equation (\ref{eq:field equations g}) is integrable \cite{BelinskyZakharov1,BelinskyVerdaguer,BelinskyZakharov2}. This means the existence of a linear system of differential equations (which have as consistency relation the equation (\ref{eq:field equations g}))
\begin{eqnarray}
\label{eq:linear system}
	\frac{d\Psi}{d\varsigma} = \frac{A(1)}{\alpha(1-\gamma)}\Psi , \,\,\,\,\,\,\,\,
	\frac{d\Psi}{d\eta} = \frac{A(-1)}{\alpha(1+\gamma)}\Psi,
\end{eqnarray}
 where we use the notation of \cite{KorotkinNicolai} because the symmetry $\varsigma\leftrightarrow\eta$ of 
(\ref{eq:field equations g}) becomes more evident in the system (\ref{eq:linear system}, \ref{eq:differentials}, \ref{eq:gamma equation general alpha} ). Reality is ensured via $\Psi^{*}(\gamma^{*})=\Psi(\gamma)$. The system (\ref{eq:linear system}) has as compatibility conditions equation (\ref{eq:field equations g}) and the zero-curvature condition \cite{FaddeevTakhtajan} associated with this integrable system \cite[p. 15, eq. 1.48]{BelinskyVerdaguer}.
The differentials are given by 
\begin{equation}
 \label{eq:differentials}
 \frac{d}{d\varsigma}=\frac{\partial}{\partial\varsigma}+\gamma_{,\varsigma}\frac{\partial}{\partial\gamma}, \,\,\,\,\,\,\,\,
 \frac{d}{d\eta}=\frac{\partial}{\partial\eta}+\gamma_{,\eta}\frac{\partial}{\partial\gamma},
\end{equation}
where $ \alpha = a(\varsigma) - b(\eta)$ is a solution of (\ref{eq:alphaequation}) and \(\gamma\) is a 'variable' spectral parameter satisfying 
\begin{equation}
\label{eq:gamma equation general alpha}
\gamma_{,\varsigma}=\frac{\alpha_{,\varsigma}}{\alpha} \frac{\gamma\left(1+\gamma\right)}{\left(1-\gamma\right)},
 \gamma_{,\eta}= \frac{\alpha_{,\eta}}{\alpha} \frac{\gamma\left(1-\gamma\right)}{\left(1+\gamma\right)},
\end{equation}
which are equivalent to 
\begin{equation}
	\label{eq:quadratic}
	\gamma+\frac{1}{\gamma}=2\frac{(w-\beta)}{\alpha}
\end{equation}
and are solved by 
\begin{equation}
\label{eq:gamma+-}
 \gamma_{\pm}(w;\varsigma,\eta) = \frac{1}{\alpha}\left\{w-\beta \pm \sqrt{(w-\beta)^{2}-\alpha^{2}}\right\} = 1/\gamma_{\mp},
\end{equation}
where $w$ complex constant and $\beta=a(\varsigma)+b(\eta)$ is a second solution of (\ref{eq:alphaequation}). The choice $\alpha=t$, $\beta=z$ corresponds to the cosmological case while $\alpha=\rho$, $\beta=t$
corresponds to cylindrically symmetric gravitational waves.

Solutions of equations (\ref{eq:linear system}) can be reproduced from a known background solution according to the B-Z dressing procedure \cite{BelinskyZakharov1, BelinskyZakharov2, BelinskyVerdaguer}: One starts with $\Psi_{0}$ a solution of (\ref{eq:linear system})(corresponding to a background metric $g_{0}$) and form
\begin{equation}
\label{eq: soliton ansatz}
 \Psi = \chi(\varsigma,\eta,\gamma)\, \Psi_{0},
\end{equation}
where 
\begin{equation}
	\label{eq:chi}
	\chi = \mbox{I}+\sum_{k=1}^{N}\frac{R_{k}(\varsigma,\eta)}{\gamma-\gamma_{k}}.
\end{equation}
where I unit matrix, and the poles $\gamma_k$, which have to be solutions of (\ref{eq:quadratic}) ie. are given by (\ref{eq:gamma+-}) for $w=w_k$, correspond to solitons in the sense understood in the inverse scattering litterature.
Reality is ensured via $\chi^{*}(\gamma^{*})=\chi(\gamma)$.

It is important that the variables $R_{k}$ have no $\gamma$ dependence and are only functions of $\varsigma$ and $\eta$. The poles $\gamma_{k}$ can be interpreted as the null trajectories of perturbations propagating on the background solution and can be thought of as gravitational solitons\cite{BelinskyVerdaguer} although they are not fully analogous to solitons as they are known in other integrable pde's \cite{KordasPhD, Kordasgravibreather, Gleiser}. The poles come in pairs either real $(\gamma_k^{+}, \gamma_k^{-})$ or $(\gamma_k, \gamma_k^{*})$ in order to ensure reality \cite{BelinskyVerdaguer, KorotkinNicolai}.

It is ensured that $g$ is symmetric via
\begin{equation}
\label{eq:chichiT}
g^{-1}\chi(\gamma) = (\chi^{-1}(1/\gamma))^T g_0^{-1}
\end{equation}

\section{The connection $A$, the Transition Matrix and observables}
\label{sec: spectral current}

Following \cite{KorotkinNicolai} we consider the Lie algebra-valued connection or logarithmic derivative of $\Psi$ $A(\gamma)$ given by 
\begin{equation}
 \label{eq:current}
 A(\gamma) = \Psi_{,\gamma}\Psi^{-1}.
\end{equation}
From (\ref{eq:differentials}, \ref{eq:current}, \ref{eq:linear system}) at $\gamma = \pm 1$ we see how the definitions (\ref{eq:A(+-1)}) arise.
Now we consider the pde's for $A(\gamma)$. Differentiating the r.h.s. of (\ref{eq:current}) w.r.t. $\eta$, $\varsigma$ and using the Lax pair (\ref{eq:linear system}) we obtain
\begin{eqnarray}
	\label{eq:A diff eq1}
A_{,\varsigma}(\gamma)=\left[ A_{+}, A(\gamma) \right] + A_{+,\gamma},\\
	\label{eq:A diff eq2}
  A_{,\eta}(\gamma)= \left[ A_{-}, A(\gamma)\right] + A_{-,\gamma},
\end{eqnarray}
where 
\begin{equation}
A_{+} = \frac{A(1)}{\alpha(1-\gamma)}-\gamma_{,\varsigma}A(\gamma)\,(= \Psi_{,\varsigma}\Psi^{-1}) , \,\,\,\,
A_{-} = \frac{A(-1)}{\alpha(1+\gamma)}-\gamma_{,\eta}A(\gamma) \,(= \Psi_{,\eta}\Psi^{-1})
\end{equation}
where we see that the equations 'decouple' for the two variables $\varsigma$ and $\eta$ a fact first observed in \cite{KorotkinNicolai-PRL}. These equations have appeared in Chapter 6 of \cite{SamtlebenThesis}. $A_{\pm}$ satisfy the zero curvature condition
\begin{equation}
\label{eq:zerocurvA+-}
A_{+,\eta}(\gamma)-A_{-,\varsigma}(\gamma) = \left[ A_{-}(\gamma), A_{+}(\gamma) \right]
\end{equation}
Also it can be shown  \cite{BelinskyZakharov1} that $A(\gamma)$ satisfies
\begin{equation}
  \label{eq:PsiPsiT}
  \Psi^{T}(\frac{1}{\gamma},\varsigma,\eta) g^{-1}(\varsigma,\eta) \Psi(\gamma,\varsigma,\eta) = g^{-1}(\varsigma,\eta),
\end{equation}
which upon differentiation gives \cite{KorotkinNicolai, SamtlebenThesis},
\begin{equation}
  \label{eq:AAT}
  \gamma^{2} A(\gamma) g = g A^{T}(1/\gamma).
\end{equation}

Also it can be seen from (\ref{eq:A(+-1)}), $A(\pm 1)$ satisfies 
\begin{equation}
  \label{eq:A(+1)=-A(-1)}
  A(\varsigma,\eta,\gamma = 1) = -A(\eta,\varsigma,\gamma=-1) ,
\end{equation}
since $\alpha = \varsigma-\eta$. This relation appears in \cite{KorotkinNicolai} as a reality condition for 
$A(\pm 1)$since for stationary axisymmetric systems $\alpha=i\rho$.
Now observing (\ref{eq:A diff eq1}), (\ref{eq:A diff eq2}) and (\ref{eq:gamma equation general alpha}) we see that the transformation ${\cal \tau}:(\varsigma, \eta, \gamma)\rightarrow(\eta, \varsigma, -\gamma)$ is an `involution' of the differential equations sending essentially the one to the other ie 
\begin{equation}
\label{eq:tau diff}
{\cal \tau}:(\frac{{\mbox d}}{{\mbox d}\varsigma}, \frac{{\mbox d}}{{\mbox d}\eta}, \gamma)\rightarrow(\frac{{\mbox d}}{{\mbox d}\eta}, \frac{{\mbox d}}{{\mbox d}\varsigma}, \gamma^{\prime}=-\gamma)
\end{equation}

It may be further noticed that the transformation 
\begin{equation}
\label{eq:involution}
{\cal \tau}:(\varsigma, \eta, \gamma)\rightarrow(\eta, \varsigma, -\gamma)
\end{equation}
 which is equivalent, for the particular choice of $\alpha=t$, to $(\alpha, \beta, \gamma)\rightarrow(-\alpha, \beta, -\gamma)$ is an involution of $A(\gamma)$ (from (\ref{eq:current}) and (\ref{eq:tau diff})), that is it satisfies 
\begin{equation}
\label{eq:tau(A)}
{\cal \tau}:A(\gamma)\rightarrow -A(-\gamma)
\end{equation}
. Having noticed (\ref{eq:A(+1)=-A(-1)}) we define the transition matrix ${\cal T}(\xi, \varsigma, \gamma)$ 
\begin{equation}
\label{eq:transition matrix}
{\cal T}(\varsigma, \eta, \xi, \gamma) \equiv -A(\varsigma, \eta, \gamma) A^{-1}(\eta, \xi, -\gamma) = A(\varsigma, \eta, \gamma) A^{-1}(\xi , \eta, \gamma)
\end{equation}
Considering ${\cal T}_{,\varsigma}$ we obtain
\begin{eqnarray}
\label{eq:transition matrix path.o.exp.}
{\cal T}(\xi, \varsigma, \gamma) = {\cal P} e^{\int_{\xi}^{\varsigma} U(\varsigma^{'}, \eta, \gamma)  d\varsigma^{'}}\\
\label{eq: U}
U = A_{,\varsigma^{'}}(\varsigma^{'}, \eta, \gamma)  A^{-1}(\varsigma^{'}, \eta, \gamma)\\
\eta={\mbox w}_2={\mbox constant}
\end{eqnarray}
where ${\cal P}$ denotes path ordered exponential.
Further considering ${\cal T}_{,\eta}$ 
we obtain 
\begin{eqnarray}
\label{eq:transition matrix eta}
{\cal T}( \xi, \varsigma,  \gamma) = V(\xi, \gamma) {\cal T}( \xi, \varsigma, \gamma) - {\cal T}(\xi,\varsigma,  \gamma) V(\varsigma, \gamma)\\
V(\varsigma, \eta, \gamma) = {A}_{,\eta}(\varsigma, \eta, \gamma)  A^{-1}(\varsigma, \eta, \gamma)
\end{eqnarray}
that is we have obtained a transition matrix analogous to the one that is very common in integrable pde's \cite{FaddeevTakhtajan, Bernard} with the null coordinate $\eta$ playing the role of time. It should be mentioned that such a matrix was lacking for the equations of gravity in the presence of two commuting Killing vectors in the connection $A$ formulation. Of course the roles of $\eta$ and $\varsigma$ can be reversed with the definition ${\cal T}^{'}=A(\varsigma, \eta, \gamma)A^{-1}(\varsigma, \vartheta, \gamma)$ which gives,
\begin{eqnarray}
\label{eq:T'}
 {\cal T^{'}}(\eta, \vartheta, \gamma) = {\cal P} e^{\int_{\eta}^{\vartheta} V(\varsigma, \eta^{'}, \gamma)  d\eta^{'}} \\
\label{eq:V}
V = A_{,\eta^{'}}A^{-1}\\
\varsigma={\mbox w}_{1}={\mbox constant}
\end{eqnarray}
 The transition matrix is extensively used in integrable pde's and in their quantization \cite{FaddeevTakhtajan, Bernard, Korepin}. It should be stressed that the poles of $\chi$ and hence $A(\gamma)$ correspond essentially to the null trajectories of the solitons and are the light cones $w_k-\beta=\pm\alpha$ \cite{BelinskyVerdaguer}. 

It is clear that for $\eta={\mbox w}_{2}$, $\gamma=-1$ from (\ref{eq:gamma+-}) with ${\mbox w}={\mbox w}_{2}$ and 
$\gamma =1$ for $\varsigma={\mbox w}_{2}$. This holds in general for appropriate expression of $\gamma$ for 
$\eta$, $\varsigma$ constant.

In Fig. $1$ we see the path, on which the transition matrix carries $A$. We define ${\cal O}$
\begin{eqnarray}
\label{eq:O}
{\cal O} = {\cal T}^{'}(\eta, \vartheta,  \varsigma={\mbox w}_{2}, \gamma=1)
{\cal T}_{B}(\varsigma, \eta, -1, 1)
 {\cal T}(\xi, \varsigma, \eta={\mbox w}_{2}, \gamma=-1) 
\end{eqnarray}
where 
\begin{equation}
\label{eq:TB}
{\cal T}_{B}(\gamma, \gamma^{'}) = A(\varsigma, \eta, \gamma)A^{-1}(\varsigma, \eta, \gamma^{'}) 
 = {\cal P}e^{\int_{\gamma}^{\gamma^{'}}A_{,\gamma}A^{-1}d\gamma}
\end{equation}
is included to ensure continuity of $\gamma$ in $A$ along the path from 
$(\xi, \eta)\rightarrow(\varsigma, \eta)\rightarrow(\varsigma, \vartheta)$in Fig. $1$. In (\ref{eq:TB}) the path can be $\gamma=e^{i\phi}$ $\phi\in[-\pi, 0]$ which corresponds, from (\ref{eq:quadratic}) to the branch cut $w\in[\beta-\alpha, \beta+\alpha]$.(It should be stressed that at no point do we attain the singularity $\alpha=0$ which corresponds to $\eta=\varsigma$ and from \ref{eq:gamma+-} to either $\gamma=0$ or $\gamma=\infty$. With $w=w_{2}$, $\alpha\cos\phi= w_{2}-\beta$).    Now we start from ${\cal O}^{2}$ 

\begin{equation}
{\cal O}^{2} = {\cal T}^{'}{\cal T}_{B}{\cal T}{\cal T}^{'}{\cal T}_{B}{\cal T}
\end{equation}
Now
\begin{eqnarray}
\label{eq:TT'}
{\cal T}^{'}{\cal T} =    {\cal P}e^{\int_{\eta}^{\vartheta}A_{,\eta^{'}}(\varsigma, \eta^{'}, 1)A^{-1}(\varsigma, \eta^{'}, 1)d\eta^{'}}{\cal P}e^{\int_{\xi}^{\varsigma}A_{,\varsigma^{'}}(\varsigma^{'}, \eta, -1)
A^{-1}(\varsigma^{'}, \eta, -1)d\varsigma^{'}}  \\ 
\nonumber
= {\cal P}e^{\int_{\eta}^{\vartheta}A_{,\eta^{'}}(\varsigma, \eta^{'}, 1)A^{-1}(\varsigma, \eta^{'}, 1)d\eta^{'}}
{\cal P}e^{\int_{\xi}^{\varsigma}A_{,\varsigma^{'}}(\eta, \varsigma^{'},1)
A^{-1}( \eta, \varsigma^{'},1)d\varsigma^{'}} = I
\end{eqnarray}
where we have used (\ref{eq:A(+1)=-A(-1)}), the particular path of Fig. $1$ which implies $\xi=\vartheta$ and 
$\eta=\varsigma$ in the boundaries of the above path integrals and the basic feature of path integrals that 
$({\cal P}e^{\int_{x}^{y}...})^{-1}={\cal P}e^{\int_{y}^{x}...}$.

Also, in a similar way 
\begin{eqnarray}
\nonumber
{\cal T}_{B}{\cal T}_{B} =  {\cal P}e^{\int_{\gamma=-1}^{\gamma^{'}=1}A_{,\gamma}(\varsigma, \eta, \gamma)
A^{-1}(\varsigma, \eta, \gamma)d\gamma}{\cal P}e^{\int_{\gamma=1}^{\gamma^{'}=-1}A_{,-\gamma}(\eta, \varsigma, -\gamma)
A^{-1}(\eta, \varsigma, -\gamma)d(-\gamma)} \\
\label{eq:TBTB}
 = A(\varsigma, \eta, 1)A^{-1}(\varsigma, \eta, -1)
A(\eta, \varsigma,  1)A^{-1}(\eta, \varsigma,-1)=I
\end{eqnarray}
where we have used ((\ref{eq:A(+1)=-A(-1)}), (\ref{eq:tau(A)}) and $\gamma(\varsigma, \eta, w)=-\gamma(\eta, \varsigma, w)$ for the case $\alpha=t$.
Since ${\cal O}^{2}=I={\cal P}e^{2\int ...}$, implies taking the square root of both sides ${\cal O}=\pm I$. Now we consider $Tr{\cal O}$, using (\ref{eq:TT'})
\begin{equation}
\label{eq:dO=0}
\frac{\partial}{\partial t}Tr{\cal T}_{B} = 
           (\frac{\partial}{\partial \varsigma}-\frac{\partial}{\partial \eta})Tr{\cal T}_{B} = 0
\end{equation}
which implies by actually performing the integration
\begin{equation}
\label{eq:obs}
\frac{\partial}{\partial t}Tr(A(\varsigma, \eta, 1)A^{-1}(\varsigma, \eta, -1)) = 0.
\end{equation}
Further it is clear that $\frac{\partial}{\partial t}Tr({\cal O}^{n})=0$	since ${\cal O}^{n}=I={\cal P}e^{n\int ..}$ which implies $\frac{\partial}{\partial t}Tr({\cal T}_{B}^{n})=0$ and hence
\begin{equation}
\label{eq:obsn}
\frac{\partial}{\partial t}Tr \left[ \left( A(\varsigma, \eta, 1) \right)^{n}\left( A^{-1}(\varsigma, \eta, -1) \right)^{n} \right] = 0
\end{equation}

\begin{figure}[h]
\begin{picture}(200,200)(0,0)
\put(0,100){\vector(1,0){220}}
\put(20,180){\vector(1,1){10}}
\put(31,185){\makebox{$\varsigma$}}
\put(20,180){\vector(-1,1){10}}
\put(4,185){\makebox{$\eta$}}
\put(20,100){\line(1,1){80}}
\put(20,100){\line(1,-1){80}}
\put(11,103){\makebox{w$_{1}$}}
\put(180,100){\line(-1,-1){80}}
\put(180,100){\line(-1,1){80}}
\put(181,103){\makebox{w$_{2}$}}
\put(105,10){\makebox{$(\xi, \eta)$}}
\put(181,92){\makebox{$(\varsigma, \eta)$}}
\put(210,103){\makebox{$\beta$}}
\put(105,181){\makebox{$(\varsigma, \vartheta)$}}
\put(0,92){\makebox{$(\xi, \vartheta)$}}
\put(10,-2){Fig.1 the path $(\xi, \eta)\rightarrow(\varsigma, \eta)\rightarrow(\varsigma, \vartheta)$}
\end{picture}
\protect\label{fig:1}
\end{figure}

In the case $\alpha=\rho, \beta=t$ which corresponds to cylindrically symmetric gravitational waves, ie. $\alpha$ spacelike, we have 
\begin{equation}
\label{eq:gammarho}
\gamma_{\pm} = \frac{1}{\rho}\left( w-t \pm \sqrt{(w-t)^{2}-\rho^{2}}\right)
\end{equation}
with $\varsigma=\frac{1}{2}(\rho+t)$, $\eta=\frac{1}{2}(\rho-t)$
In this case the necessary involution is $(\varsigma\leftrightarrow\eta)$ (which corresponds to 
$t\rightarrow-t$) along with $w\rightarrow-w$ which corresponds to $\gamma_{+}(\rho, t, w)\rightarrow\gamma_{+}(\rho, -t,-w)=-\gamma_{-}(\rho, t, w)=-1/\gamma_{+}(\rho, t, w)$. It is again evident from (\ref{eq:differentials}) that 
the involution $\tau^{'}$ defined by
\begin{equation}
\label{eq:tau'}
{\cal \tau}^{'}:(\varsigma, \eta, w)\rightarrow(\eta, \varsigma, -w)
\end{equation}
has the effect $\tau^{'}:A(\varsigma, \eta, \gamma)\rightarrow \gamma_{-}^{2} A(\eta, \varsigma, -\gamma_{-})$
and $A(\gamma_{-})$ satisfies the same differential equation as $A(\gamma_{+})$ (because (\ref{eq:gamma+-}) are the two solutions of (\ref{eq:quadratic})).
Hence $\tau^{'}$ is an involution of the linear system (\ref{eq:linear system})
\begin{equation}
\label{eq:tau' diff}
{\cal \tau}^{'}:(\frac{{\mbox d}}{{\mbox d}\varsigma}, \frac{{\mbox d}}{{\mbox d}\eta}, \gamma_{+})\rightarrow(\frac{{\mbox d}}{{\mbox d}\eta}, \frac{{\mbox d}}{{\mbox d}\varsigma}, \gamma^{\prime}=-\gamma_{-})
\end{equation}
Hence transition matrices can be defined for the case $\alpha=\rho$ (ie. $\alpha$ spacelike) as
\begin{eqnarray}
\label{eq:Troman}
{\mbox{\rmfamily T}} = A(\xi, \eta, \gamma_{+})A^{-1}(\eta, \varsigma, -\gamma_{-}) = 
A(\xi, \eta, \gamma_{+})A^{-1}(\varsigma, \eta, \gamma_{+}) \\
\label{eq:Troman'}
{\mbox{\rmfamily T}}^{'} = A(\varsigma, \eta, \gamma_{+})A^{-1}(\vartheta, \varsigma, -\gamma_{-}) = 
A(\varsigma, \eta, \gamma_{+})A^{-1}(\varsigma, \vartheta, \gamma_{+})
\end{eqnarray}
Hence we have obtained transition matrices for $\alpha$ spacelike in a similar way to the timelike case above. (In this and the next chapter all commutators are matrix commutators.)

The metric (\ref{eq:metric}) corresponds to a wide variety of solutions including cosmological, cylindrically symmetric gravitational waves, and stationary axisymmetric space-times (with appropriate transcription of the coordinates). The solitonic ansatz, among other methods, may be employed to obtain solutions (from a diagonal background usually but not exclusively) from a seed solution including Schwarzschild and Kerr among many others. Although the considerations here involve mainly space-times with one of the two significant coordinates timelike, the observables may be relevant for the axistationary case, e.g. inside a Black Hole horizon where one of the two significant coordinates becomes timelike. 

A generic diagonal seed spacetime for the solitonic technique \cite{BelinskyVerdaguer} is 
\begin{eqnarray}
 \label{eq:g0}
 (g_{0})_{11} =  \alpha e^{\beta}   \\
 (g_{0})_{22} =  \alpha e^{-\beta}  
\end{eqnarray}
The corresponding solution of (\ref{eq:linear system}) is given by 
\begin{eqnarray}
 \label{eq:Psi0}
 (\Psi_{0})_{11} =  \alpha(\gamma^{2}+2\frac{\beta}{\alpha}\gamma+1)^{1/2} e^{\frac{1}{2}\alpha\gamma+\beta}  \\
 (\Psi_{0})_{22} =  \alpha(\gamma^{2}+2\frac{\beta}{\alpha}\gamma+1)^{1/2} e^{-\frac{1}{2}\alpha\gamma-\beta}   
\end{eqnarray}
It is straightforward to check that (e.g. with $\alpha=\varsigma-\eta$, $\beta=\varsigma+\eta$), $(\frac{\partial}{\partial\varsigma}-\frac{\partial}{\partial\eta})Tr(g_{0,\varsigma}g_{0}^{-1}g_{0,\eta}g_{0}^{-1})=0$.
It is well-known \cite{BrMaiGib} that the equations (\ref{eq:field equations g}) are also valid for $d$-dimensional general relativity in the presence of $d-2$ Killing vectors. The considerations here are valid in that case also and the observables are available to use in string-theoretic considerations.

\section{The curvature B and observables}
\label{sec:B}
We now consider 
\begin{equation}
\label{eq:B}
  B=A_{,\gamma}A^{-1}
\end{equation}
appearing in (\ref{eq:TB}).
Taking the derivative of $B$ with respect to $\varsigma$, $\eta$ turning the partial derivatives on the rhs of 
(\ref{eq:B}) and using (\ref{eq:A diff eq1}, \ref{eq:A diff eq2}) one obtains
\begin{eqnarray}
\label{eq:Bzeta}
B_{,\varsigma} = \left[ B_{+} , B \right] + B_{+,\gamma }   \\
\label{eq:Beta}
B_{,\eta} = \left[ B_{-} , B \right] + B_{-,\gamma}
\end{eqnarray}
where
\begin{eqnarray}
\label{eq:B+}
B_{+} = \left( \left[ A_{+}, A \right] + A_{+,\gamma} \right) A^{-1} \,\,\,\,\, (=A_{,\varsigma}A^{-1}) \\
\label{eq:B-}
B_{-} = \left( \left[ A_{-}, A \right] + A_{-,\gamma} \right) A^{-1} \,\,\,\,\, (=A_{,\eta}A^{-1})
\end{eqnarray}
From (\ref{eq:A diff eq1}, \ref{eq:A diff eq2}) we see that
\begin{eqnarray}
\nonumber
\lefteqn{B_{+}(\varsigma, \eta, -1+\delta\gamma)} \\
\nonumber
&=& A_{,\varsigma}(\varsigma, \eta, -1+\delta\gamma)A^{-1}(\varsigma, \eta, -1+\delta\gamma) \\
\nonumber
&=&\Bigl( \left[ \frac{A(1)}{\alpha(1-\gamma)}, A(-1+\delta\gamma) \right] A^{-1}(-1+\delta\gamma) \\
& &+ \frac{A(1)A^{-1}(-1+\delta\gamma)}{\alpha(1-\gamma)^{2}}-(\gamma_{,\varsigma})_{,\gamma} I-\gamma_{,\varsigma}A_{,\gamma}(-1+\delta\gamma)A^{-1}(-1+\delta\gamma)\Bigr)
\label{eq:A(-1)zeta}
\end{eqnarray} 
Also
\begin{eqnarray}
\nonumber
\lefteqn{B_{-}(\varsigma, \eta, 1+\delta\gamma) =}\\
&& A_{,\eta}(1+\delta\gamma)A^{-1}(1+\delta\gamma) =\nonumber\\
\nonumber
&&\left[ \frac{A(-1)}{2\alpha}, A(1+\delta\gamma) 
\right] A^{-1}(1+\delta\gamma) + \frac{A(-1)A^{-1}(1+\delta\gamma)}{4\alpha} - 
\frac{I}{2\alpha} \\
&&-\frac{\delta\gamma}{2\alpha}A_{,\gamma}(1+\delta\gamma)A^{-1}(1+\delta\gamma) \label{eq:A(1)eta}
\end{eqnarray}
We form,
\begin{eqnarray}
\frac{2\alpha}{\delta\gamma}A_{,\eta}(1+\delta\gamma)A^{-1}(1+\delta\gamma)
-\frac{2\alpha}{\delta\gamma}A_{,\eta}(1)A^{-1}(1) 
- 2\alpha \Bigl\langle\bigl(A_{,\eta}(\gamma)A^{-1}(\gamma)\bigr)_{,\gamma}\Bigr\rangle_{\gamma=1} = 0
\end{eqnarray}
and hence, using (\ref{eq:A diff eq2}, \ref{eq:A diff eq1}, \ref{eq:A(1)eta}, \ref{eq:A(-1)zeta})
\begin{eqnarray}
B(1) = \Bigl[\frac{1}{2}A(1), A(-1)\Bigr]A^{-1}(1)+\frac{1}{2}A(-1)A^{-1}(1) - I
\end{eqnarray}
and,
\begin{eqnarray}
B(-1) =  \Bigl[\frac{1}{2}A(-1), A(1) \Bigr] A^{-1}(-1) -\frac{1}{2}A(1)A^{-1}(-1) + I 
\end{eqnarray}

Further we want to calculate the limit $B_{+}(\gamma\rightarrow 1)$ and $+B_{-}(\gamma\rightarrow -1)$.
From (\ref{eq:B+}, \ref{eq:B-}) in the limit $\gamma\rightarrow 1$ we get
\begin{eqnarray}
\nonumber
\lefteqn{A_{,\varsigma}(\varsigma, \eta, 1)A^{-1}(1)=B_{+}(\varsigma, \eta, \gamma\rightarrow 1)=} \\
\label{eq:B+(+1)app}
& & \left[ -A(1), \frac{B(1)}{\alpha} \right]  +  \frac{\bigl\langle B_{,\gamma}\bigr\rangle _{\gamma=1}}{\alpha}\\
\lefteqn{A_{,\eta}(\varsigma, \eta, -1)A^{-1}(-1)=B_{-}(\varsigma, \eta, \gamma\rightarrow -1)=} \\
\label{eq:B-(-1)app}
& & \left[ A(-1), \frac{B(-1)}{\alpha}\right]  +    \frac{\bigl\langle B_{,\gamma}\bigr\rangle _{\gamma=-1}}{\alpha} 
\end{eqnarray}
where we have used 
\begin{eqnarray}
\bigl.\gamma_{,\varsigma}\bigr|_{(1+\delta\gamma)}\approx -\frac{2}{\alpha\delta\gamma}\approx 
\bigl.-\gamma_{,\eta}\bigr|_{(-1+\delta\gamma)}    \\
\bigl.(\gamma_{,\varsigma})_{,\gamma}\bigr|_{(1+\delta\gamma)}\approx \frac{2}{\alpha\delta\gamma^{2}}\approx \bigl.-(\gamma_{,\eta})_{,\gamma}\bigr|_{(-1+\delta\gamma)} \\
\frac{\delta A}{\delta\gamma}A^{-1}\frac{\delta(A^{-1})^{-1}}{\delta\gamma}A^{-1} = -I
\end{eqnarray}

We see from (\ref{eq:tau(A)}, \ref{eq:B}) (in the case $\alpha=t$ ie. $\alpha$ timelike)
\begin{eqnarray}
\label{eq:tau(B+)}
{\tau}:B_{+}(\varsigma, \eta, \gamma) \rightarrow B_{-}(\eta, \varsigma, -\gamma) \\
\label{eq:tau(B)}
{\tau}:B(\varsigma, \eta, \gamma) \rightarrow -B(\eta, \varsigma, -\gamma)
\end{eqnarray}
The transformation $\tau$ is again an involution of the equations (\ref{eq:Bzeta}, \ref{eq:Beta}) so like in section \ref{sec: spectral current} we can define transition matrices for $B$. The process of obtaining observables this way must involve the derivatives of $A_{\pm}, B_{\pm}$ and
commutators thereof, always modulo the integrable systems zero curvature condition, the field equations 
(\ref{eq:field equations g}) and Bianchi identity.  

So we have (for $\alpha=t$ timelike)
\begin{eqnarray}
\label{eq:Tcurv} 
{\mbox \ttfamily T} = -B(\varsigma, \eta, \gamma)B^{-1}(\eta, \xi, -\gamma) = {\cal P}e^{\int_{\xi}^{\varsigma}B_{,\varsigma^{'}}(\varsigma^{'}, \eta, \gamma)  B^{-1}(\varsigma^{'}, \eta, \gamma)d\varsigma^{'}}
\\
\label{eq:T'curv}
{\mbox \ttfamily T}^{'} = -B(\varsigma, \vartheta, \gamma)B^{-1}(\eta, \varsigma, -\gamma) = {\cal P}e^{\int_{\eta}^{\vartheta}B_{,\eta^{'}}(\varsigma, \eta^{'}, \gamma)  B^{-1}(\varsigma, \eta^{'}, \gamma)d\eta^{'}}
\\ \label{eq:TC}
{\mbox \ttfamily T}_{C}(\gamma, \gamma^{'}) = B(\varsigma, \eta, \gamma)B^{-1}(\varsigma, \eta, \gamma^{'}) 
 = {\cal P}e^{\int_{\gamma}^{\gamma^{'}}B_{,\gamma}B^{-1}d\gamma}
\end{eqnarray}
and the observable
\begin{equation}
{\mbox Tr}{\cal O}_{B} = {\mbox Tr}\left( {\mbox \ttfamily T}^{'}{\mbox \ttfamily T}_{C} {\mbox \ttfamily T} \right)
\end{equation}
which gives 
\begin{equation}
\label{eq:Bobs}
\left(\frac{\partial}{\partial\varsigma}-\frac{\partial}{\partial\eta}\right){\mbox Tr}\Bigl(B^{n}(1)B^{n}(-1)\Bigr)
\end{equation}
since
\begin{equation}
{\cal O}_{B} B(\xi, \eta, \gamma) =  B(\varsigma, \vartheta, -\gamma)
\end{equation}
and with the boundary condition (\ref{eq:tau(B)}) on the path of Fig. $1$ we get (\ref{eq:Bobs}).

We may consider obtaining observables directly from ${\mbox Tr}{\cal O}$. Indeed
\begin{eqnarray}
\nonumber
\lefteqn{\left( \frac{\partial}{\partial\varsigma}-\frac{\partial}{\partial\eta} \right) {\mbox Tr}{\cal O}} \\ \nonumber
&=&\left( \frac{\partial}{\partial\varsigma}-\frac{\partial}{\partial\eta} \right) {\mbox Tr}
\left( {\cal T}^{'}{\cal T}_{B}{\cal T} \right) \\
\label{eq:dTrT'TBT}
&=& \left( \frac{\partial}{\partial\varsigma}-\frac{\partial}{\partial\eta} \right) \left( {\cal P}e^{\int_{\eta}^{\vartheta}B_{-}(\varsigma, \eta^{'}, 1)
d\eta^{'}}{\cal P}e^{\int_{-1}^{1}B(\varsigma, \eta, \gamma)d\gamma}{\cal P}e^{\int_{\xi}^{\varsigma}B_{+}(\varsigma^{'}, \eta, -1)d\varsigma^{'}}   \right)
\label{eq:dO=0details}
\end{eqnarray}
Taking the derivative in (\ref{eq:dTrT'TBT}), and using (\ref{eq:TT'}), we have
\begin{eqnarray}
\nonumber
\lefteqn{{\mbox Tr}{\cal O} =} \\
\nonumber
&&{\mbox Tr}\left( \left( \int_{\eta}^{\vartheta}B_{-,\varsigma}(\varsigma, \eta^{'}, 1)d\eta^{'} \right) {\cal T}_{B}-{\cal T}_{B} \left( \int_{\xi}^{\varsigma}B_{+,\eta}(\varsigma^{'}, \eta, -1)d\varsigma^{'} \right) \right.+
\\
&& \Bigl. {\cal T}_{B}B_{+}(\varsigma, \eta, -1)-B_{-}(\varsigma, \eta, +1){\cal T}_{B} + {\cal T}_{B,\varsigma}-{\cal T}_{B,\eta} \Bigr) 
\label{eq:obscurv}
\\
&=& 0
\end{eqnarray}

So we have obtained in (\ref{eq:obscurv}) and  a relation giving observables in terms of the curvature variables $B_{\pm}(\mp1)$ $B(\gamma)$ in what is a form of generalised zero curvature condition.
It is also clear that there exists a countable infinite hierarchy of hierarchies of constants of motion built from $A(\gamma), B(\gamma), ...$.

\section{Field equations and interpretation of the connection observable}
\label{sec:discussion}
First consider (\ref{eq:obs}). Using the fact that the inverse of a $2\times 2$ matrix $M$ can be written as
\begin{equation}
\label{eq:2x2matrTrdet}
M^{-1} = \frac{{\mbox Tr}M \, I-M}{det M}
\end{equation}
(\ref{eq:obs}) may be written as
\begin{equation}
{\mbox Tr} \bigl( A(1)A^{-1}(-1) \bigr) = \frac{{\mbox Tr}A(1){\mbox Tr}A(-1)-{\mbox Tr}\bigl(A(1)A(-1)\bigr)}{\det A(-1)}
\end{equation}
Notice immediately that from (\ref{eq:obs}) the alpha dependence cancels. So the observables obtained here are well behaved, in the first instance, on the axis 
$\alpha=0$, which for $\alpha$ timelike is the cosmological singularity and in the cylindrically symmetric case is the axis of symmetry.
Further we may traverse, in Fig. \ref{fig:1}, the path in the opposite direction 
$(\varsigma, \vartheta)\rightarrow(\varsigma, \eta)\rightarrow(\xi, \eta)$, in which case everything in Sec. 
\ref{sec: spectral current} may be repeated to obtain the observable 
\(Tr\left(A(-1)A^{-1}(1)\right)\)
This implies that
\begin{equation}
\label{eq:obsdet}
\det\left(A(1)A^{-1}(-1)\right)
\end{equation}
is a constant of motion.
This suggests that the first essential constant of motion is Tr$(A(1)A(-1))$. Indeed if we consider
\begin{eqnarray}
\label{eq:obssimple} 
\left[\partial_{\varsigma}-\partial_{\eta}\right]
{\mbox Tr}\left(g_{,\varsigma}g^{-1}g_{,\eta}g^{-1}\right) = F(\varsigma, \eta) - F(\eta, \varsigma) = E(\varsigma, \eta)\,\,\,{\mbox where,}\\
\,\,\,
F(\varsigma, \eta) = Tr\left(g_{,\varsigma\varsigma}g^{-1}g_{,\eta}g^{-1}
+ g_{,\eta\varsigma}g^{-1}g_{,\eta}g^{-1} - 2 g_{,\varsigma}g^{-1}g_{,\varsigma}g^{-1}g_{,\eta}g^{-1} \right)
\end{eqnarray}

and we have used (\ref{eq:A(+1)=-A(-1)}). 
Upon $\varsigma\leftrightarrow\eta$ (which corresponds to $t\rightarrow -t$) 
$E(\varsigma, \eta)=-E(\varsigma, \eta)$ hence $E(\varsigma, \eta)=0$ and so $Tr(A(1)A(-1))$ is a constant of motion. There has been indication \cite{Belinskybr} that $Tr(A(1)A(-1))$ is a constant of motion.
Of course upon quantization if $A, B\in$ su($2$) or so(2) we have \(Tr(AB^{-1})=Tr(AB)\).

The Einstein-Hilbert action is \cite{Thiemann}
\begin{equation}
\label{eq:action}
 {\cal S} = \int d^{4}x \sqrt{\left|\det g_{\mu\nu}\right|}\, R^{(4)}
\end{equation}

Using \cite{Thiemann},

\begin{equation}
\label{eq:Rextrinsic}
{\cal R}^{(D+1)} = {\cal R}^{(D)} - s\left[K_{\mu\nu}K^{\mu\nu}-(K_{\mu}^{\,\,\mu})^2\right]
                    +2s\nabla_{\mu}\left(n^{\nu}K_{\nu}^{\,\,\mu}-n^{\mu}K_{\nu}^{\,\,\nu}\right)
\end{equation}
twice, where $K_{\mu\nu}$ is the extrinsic curvature, $s$ is the signature of the metric and $n^{\mu} = \frac{1}{N} \frac{\partial}{\partial t}$ is the normal of the $D$-surface for $D=3$, $n^{\mu} = \frac{1}{N} \frac{\partial}{\partial z}$ for $D=2$ (we view the two hypersurface orthogonal Killing vectors metric (\ref{eq:metric}) as $1-1-2$ metric and apply (\ref{eq:Rextrinsic}) twice) we obtain

\begin{eqnarray}
\label{eq:R4}
{\mathcal L}& =  &\frac{\alpha}{4} g_{ac,_{t}}g^{cd}g_{de,_{t}}g^{ea}
           - \frac{\alpha}{4} g_{ac,_{z}}g^{cd}g_{de,_{z}}g^{ea} \nonumber
			\\		&    &-2\alpha_{,t}\sigma_{,t} 
					+ 2\alpha_{,z}\sigma_{,z} - \frac{\alpha_{,t}^{\,\,\, 2}}{\alpha} + \frac{\alpha_{,z}^{\,\,\, 2}}{\alpha} - 2 \alpha_{,zz}  - 2\alpha N^{2} \partial_{t}\left(\frac{\sigma_{,t}}{N^{2}}+\frac{\alpha_{,t}}{\alpha N^{2}}\right)        
\end{eqnarray}
where the last term is a boundary term. 
The Euler-Lagrange equations for the metric functions $g_{ab}(z,t)$ give
\begin{equation}
\label{eq:field equations z-t}
\left(g_{,z}g^{-1}\right)_{,z}-\left(g_{,t}g^{-1}\right)_{,t} = 0
\end{equation}
which is exactly (\ref{eq:field equations g}) in the variables $(z,t)$. The trace of 
(\ref{eq:field equations z-t}) gives
\begin{equation}
\label{eq:field equations alpha z-t}
\alpha_{,zz} = \alpha_{,tt}
\end{equation}
ie (\ref{eq:field equations alpha}). If $\alpha = a(\varsigma) - b(\eta)$ is a timelike solution of (\ref{eq:field equations alpha}) then $\beta = a(\varsigma) + b(\eta)$ is an a second independent spacelike solution of 
(\ref{eq:field equations alpha}, \ref{eq:field equations alpha z-t}). The Euler-Lagrange equations give the equation for $\sigma$
\begin{equation}
\label{eq:field equation sigma}
\sigma_{,zz}-\sigma_{,tt}= \frac{1}{8} Tr\Bigl(  \left(g_{,t}g^{-1}\right)^{2} - 
\left(g_{,z}g^{-1}\right)^{2} \Bigr)
\end{equation}
It is clear from the above discussion that the right hand side of 
(\ref{eq:field equation sigma}) is a constant. Further we see that $\alpha_{,z} = a^{'} - b^{'} = \beta_{,t}$, $\beta$ spacelike. Since $N$ is the lapse that is it measures proper time and also proper distance in the $z$ direction we interpret (\ref{eq:field equation sigma}) as a dispersion relation that is an equation of the form $E^{2}-p^{2}=m^2$. That means that our connection observable is interpreted as a constant that corresponds to mass. This makes equation (\ref{eq:field equation sigma}) a typical Klein-Gordon equation and makes available all the tools of the corresponding field theory for quantization. 
With the definite choice of variables $\alpha = t$, $\beta = z$ equation 
(\ref{eq:field equation sigma}) is solved by \cite{BelinskyVerdaguer} (transcribed to the variables used here)
\begin{equation}
\label{eq:sigmat sigmaz}
\sigma_{,t} = \frac{\alpha}{8}Tr\Bigl(\left(g_{,t}g^{-1}\right)^{2} 
+ \left(g_{,z}g^{-1}\right)^{2} \Bigr) + C_{,t} \, , 
\, \sigma_{,z} = \frac{\alpha}{4}Tr\Bigl(g_{,t}g^{-1}g_{z}g^{-1}\Bigr) + C_{,z}
\end{equation}
up to essentially a solution of the wave equation (\ref{eq:field equations alpha z-t}) $C(z,t)$.
To stress the interpretation of (\ref{eq:field equation sigma}), with variables and conjugate momenta to be $(\alpha, p_{\alpha})$,
$(\beta, p_{\beta})$, $(g, p_{g})$ where 
\begin{eqnarray}
p_{\alpha} = 2 (\sigma_{,t} + \frac{\alpha_{,t}}{\alpha}) \,\,\, , \,\,\,
p_{\beta} =  2 (\sigma_{,z} + \frac{\beta_{,t}}{\alpha}) \,\, , \,\,
p_{g}^{ab} = \frac{\alpha}{2} g^{ac}g_{cd_{,t}}g^{db}
\end{eqnarray}
we have for the Hamiltonian density ${\cal H}$
\begin{equation}
{\cal H} = p_{\mu} \dot{q}^{\mu} - {\cal L} 
\end{equation}
which gives 
\begin{equation}
{\cal H} = \frac{\alpha}{4} Tr\Bigl(\left(g_{,t}g^{-1}\right)^{2} 
+ \left(g_{,z}g^{-1}\right)^{2} \Bigr)
\end{equation}
after fixing the freedom in (\ref{eq:sigmat sigmaz}) to $C_{,t}=-\frac{1}{\alpha}$. So the energy of the system is 
$p_{t}$ and is positive definite.
Further it is clear that upon quantization $(q, p)\rightarrow(q, \frac{\partial}{\partial q})$
so the generators of $\sigma_{,t}$, $\sigma_{,z}$ are $i\frac{\partial}{\partial t}$, $i\frac{\partial}{\partial z}$ which clearly commute by construction here.
The other translation generators are trivial.
Rotation generators are
\begin{equation}
J_{1} = -i\left(x_{2}\partial_{z} - z \partial_{2} \right) \, , \, J_{2} = -i \left(z\partial_{1}
-x_{1}\partial_{z} \right) \, , 
\, J_{z} = -i \left( x_{1} \partial_{2} 
- x_{2} \partial_{1} \right)
\end{equation}
Boost generators are 
\begin{eqnarray}
K_{1} = i \left( t \partial_{1} + x_{1} \partial_{t}\right) \, , \,
K_{2} = i \left( t \partial_{2} + x_{2} \partial_{t}\right) \, , \,
K_{z} = i \left( t \partial_{z} + z \partial_{t} \right)
\end{eqnarray}
We form \cite{Ryder}
\begin{eqnarray}
L_{1} = K_{1}-J_{2} = i \left( (t+z)\partial_{1} + x_{1}\partial_{t-z} \right), \\
L_{2} = K_{2}+J_{1} = i \left( (t+z)\partial_{2} + x_{2}\partial_{t-z} \right)
\end{eqnarray}
This gives
\begin{eqnarray}
\left[K_{1}, J_{2}\right] = K_{z}, \\
\left[K_{1}, J_{z}\right] = K_{2} \\
\left[K_{2}, J_{1} \right] = -K_{z}, \\
\left[K_{1}, K_{2} \right] = -iJ_{z}, \\
\left[K_{i}, J_{i} \right] = 0 (no summation),\\
\left[L_{1}, L_{2} \right] = 0 ,\\
\left[J_{3}, L_{1} \right] = i L_{2} ,\\
\left[L_{2}, J_{3} \right] = i L_{1}
\end{eqnarray}
The last three relations represent the Euclidean group in two dimensions ISO(2).

\section{Conclusion}
\label{sec:Conclusion}

We have obtained transition matrices (\ref{eq:transition matrix path.o.exp.}) in terms of connection variables satisfying equations similar to the ones satisfied by the transition matrices for other integrable pde's. Using these, hierarchies of observables in terms of connection and curvature variables have been obtained. 
The first in the hierarchy of connection observables has been shown to correspond to the mass of a Klein-Gordon equation that governs the lapse variable $N$.

\section{Data accessibility}
Not applicable

\section{Competing interests}
We have no competing interests

\section{Authors' contributions}
Not applicable

\section{Acknowledgments}

\section{Funding statement}
Self-funded

\section{Ethics statement}
Not applicable

\bibliography{bibfile}

\begin{thebibliography}{10}

\bibitem{Geroch}
Geroch R.J.
\newblock A method for generating new solutions of einstein's equation.
\newblock {\em J. Math. Phys.}, 13:394, 1972.

\bibitem{BelinskyZakharov1}
{Belinsky V.A., Zakharov V.E.}
\newblock Integration of the einstein equations by means of the inverse
  scattering problem technique and construction of exact soliton solutions.
\newblock {\em Sov. Phys.-JETP}, 48(6):985, 1978.

\bibitem{Maison}
Maison D.
\newblock On the complete integrability of the stationary, axially symmetric
  einstein equations.
\newblock {\em J. Math. Phys.}, 20(5):871, 1979.

\bibitem{Neugebauer}
Neugebauer G.
\newblock B$\ddot{a}$cklund transformations of axially symmetric stationary
  gravitational fields.
\newblock {\em J. Phys. A: Math. Gen.}, 12:L67, 1979.

\bibitem{HauserErnst}
{Hauser I., Ernst F.J.}
\newblock Proof of a geroch conjecture.
\newblock {\em J. Math. Phys.}, 22(5):1051, 1981.

\bibitem{Harrison}
Harrison B.K.
\newblock {\em Phys. Rev. Lett.}, 41:1197, 1978.

\bibitem{HKX}
{Hoenselaers C., Kinnersley W., Xanthopoulos B.C.}
\newblock Symmetries of the stationary einstein-maxwell equations. vi.
  transformations which generate asymptotically flat spacetimes with arbitrary
  multipole moments.
\newblock {\em J. Math. Phys.}, 20:2530, 1979.

\bibitem{BelinskyVerdaguer}
{Belinski V.A., Verdaguer E.}
\newblock {\em Gravitational Solitons}.
\newblock Cambridge University Press, 2001.

\bibitem{KorotkinNicolai}
{Korotkin D., Nicolai H.}
\newblock Isomonodromic quantization of dimensionally reduced gravity.
\newblock {\em Nucl. Phys.}, B475:379, 1996.

\bibitem{kordasarx}
Kordas P.
\newblock Transition matrix, poisson bracket for gravisolitons in the dressing
  formalism.
\newblock {\em arXiv preprint arXiv:1002.0524}, 2010.

\bibitem{Rovelli}
{Rovelli C.}
\newblock {\em Quantum Gravity}.
\newblock Cambridge University Press, 2004.

\bibitem{Ashtekar}
{Ashtekar A.}
\newblock {\em Lectures on non-perturbative Canonical Gravity}.
\newblock World Scientific, 1991.

\bibitem{ExactSolutionsBook}
{Stephani H., Kramer D., MacCallum M., Hoenselaers C., Herlt E.}
\newblock {\em Exact Solutions of Einstein's equations}.
\newblock Cambridge University Press, 2003.

\bibitem{BelinskyZakharov2}
{Belinsky V.A., Zakharov V.E.}
\newblock Stationary gravitational solitons with axial symmetry.
\newblock {\em Sov. Phys.-JETP}, 50:1, 1979.

\bibitem{FaddeevTakhtajan}
{Faddeev L.D., Takhtajan L.A.}
\newblock {\em Hamiltonian Methods in the Theory of Solitons}.
\newblock Springer, 1987.

\bibitem{KordasPhD}
Kordas P.
\newblock {\em Solutions of Einstein's equations with two commuting Killing
  vectors}.
\newblock PhD thesis, QMUL, 1995.

\bibitem{Kordasgravibreather}
{Kordas P.}
\newblock Properties of the gravibreather.
\newblock {\em Phys. Rev D}, 48(10):5013, 1993.

\bibitem{Gleiser}
{Gleiser R.J., Garate A., Nicasio C.O.}
\newblock Topological properties of single gravisolitons.
\newblock {\em J. Math. Phys.}, 37:5652, 1996.

\bibitem{KorotkinNicolai-PRL}
{Korotkin D., Nicolai H.}
\newblock Separation of variables and hamiltonian formulation for the ernst
  equation.
\newblock {\em Phys. Rev. Lett.}, 74:1272--1275, 1995.

\bibitem{SamtlebenThesis}
{Samtleben H.}
\newblock {\em Classical and Quantum Symmetries in Models of Dimensionally
  Reduced Gravity}.
\newblock PhD thesis, University of Hamburg, 1998.

\bibitem{Bernard}
{Babelon O., Bernard D., Talon M.}
\newblock {\em Introduction to classical Integrable systems}.
\newblock Cambridge University Press, 2006.

\bibitem{Korepin}
{Korepin, Bogoliubov, Izergin}.
\newblock {\em Quantum Inverse Scattering Method and Correlation Functions}.
\newblock Cambridge University Press, 1997.

\bibitem{BrMaiGib}
Gibbons~G. Breitenlohner~P., Maison~D.
\newblock Four-dimensional black holes from kaluza-klein theories.
\newblock {\em Commun. Math. Phys.}, 120:295--333, 1988.

\bibitem{Belinskybr}
Belinsky V.A.
\newblock Gravitational breather and topological properties of gravisolitons.
\newblock {\em Phys. Rev. D}, 44:3109, 1991.

\bibitem{Thiemann}
{Thiemann T.}
\newblock {\em Modern Canonical Quantum General Relativity}.
\newblock Cambridge University Press, 2008.

\bibitem{Ryder}
Ryder L.H.
\newblock {\em Quantum Field Theory}.
\newblock Cambridge, 1987.

\end{thebibliography}
\bibliographystyle{unsrt}

\end{document}